\documentclass{aa} 
\usepackage[varg]{txfonts}
 
\usepackage{amsmath}
\usepackage{amsfonts}
\usepackage{amssymb}
\usepackage{graphicx}
\usepackage{fourier}

\usepackage{natbib}
\bibpunct{(}{)}{;}{a}{}{,} 

\usepackage{booktabs} 

\usepackage{makecell}
\usepackage{txfonts}
\usepackage{graphicx,natbib,url,twoopt}
\usepackage{lipsum}
\usepackage{multicol}
\usepackage{float}

\usepackage{threeparttable}
\usepackage{adjustbox}

\usepackage{caption}
\usepackage{subcaption}

\def\m2s2{\hbox{\,m$^{2}$\,s$^{-2}$}} 

\begin{document}
\title{Global dynamics and architecture of the Kepler-444 system}

\author{M. Stalport
\and E. C. Matthews
\and V. Bourrier
\and A. Leleu
\and J.-B. Delisle
\and S. Udry
} 

\institute{D\'epartement d'Astronomie, Universit\'e de Gen\`eve, Chemin Pegasi 51b, CH-1290 Versoix, Suisse \label{i:geneva}}
 
\date{Received ?? / Accepted ??} 
 
\abstract
{S-type planets, which orbit one component of multiple-star systems, place strong constraints on the planet formation and evolution models. 
A notable case study is Kepler-444, a triple-star system whose primary is orbited by five planets smaller than Venus in a compact configuration, and for which the stellar binary companion revolves around the primary on a highly eccentric orbit.} 
{Several open questions remain about the formation and evolution of Kepler-444. Having access to the most precise up-to-date masses and orbital parameters is highly valuable to tackle those questions. We provide the first full dynamical exploration of this system, with the goal to refine those parameters.}
{We apply orbital stability arguments to refine the system parameters, both on models without and with the stellar binary companion in order to understand the origin of the dynamical constraints. This approach makes use of the Numerical Analysis of Fundamental Frequencies (NAFF) fast chaos indicator. We also explore potential two and three-planet mean-motion resonances (MMR) in the system. Prior to investigating the dynamics of a model including the binary companion, we update its orbital parameters and mass using new observational constraints from both HIRES radial velocity and Gaia astrometric data, as well as archival imaging of the system.}
{The planetary system does not appear in any of low-order two or three-planet MMR. We provide the most precise up-to-date dynamical parameters for the planets and the stellar binary companion. The orbit of the latter is constrained by the new observations, and also by the stability analysis. This update further challenges the planets formation processes. 
We also test the dynamical plausibility of a sixth planet in the system, following hints observed in the Hubble Space Telescope (HST) data. We find that this putative planet could exist over a broad range of masses, and with an orbital period roughly comprised between 12 and 20 days.}
{We note an overall good agreement of the system with short-term orbital stability. This suggests that a diverse range of planetary system architectures could be found in multiple-star systems, potentially further challenging the planet formation models.}


\maketitle

\section{Introduction} 
\label{Section:Kepler-444}
Multiple-star systems represent about half of the K to F type stars of the Milky Way \citep[e.g.][]{Duquennoy1991, Offner2022}. Yet, discovered planets in these systems constitute a small minority of the entire exoplanet population known to date. 
However, this poorly studied planet population contains important constraints for planet formation and orbital evolution processes. From the small sample available, it is therefore important to ensure the highest precision on the planetary orbital parameters and masses, in order to add the strongest constraints on the models. In a former paper, we introduced a technique to rigorously refine the planetary parameters in single-star systems through orbital stability arguments \citep{Stalport2022}. This technique is based on a global parameter space exploration, from which the orbital stability is estimated on each system's configuration making up the posterior distribution. The stability is computed from the Numerical Analysis of Fundamental Frequencies fast chaos indicator \citep[NAFF,][]{Laskar1990, Laskar1993}, which we calibrate for the system under study. This approach proved efficient in refining the orbital eccentricities, particularly sensitive to the orbital stability. Furthermore, in some cases the planetary masses experience a significant update under the stability constraint. 

The present paper extends the scope of the technique to multiple-star systems. In particular, we apply it to a notable triple-star planet-hosting system: Kepler-444 \citep{Campante2015}. In this system, a primary star (Kepler-444\,A) is orbited by a tight binary star (Kepler-444\,BC). The Kepler mission revealed the existence of five s-type planets smaller than Venus orbiting the primary. They orbit in a compact configuration, with orbital periods comprised between 3.5 and 10 days. The stellar doublet Kepler-444\,BC orbits the primary in a highly eccentric orbit, hence posing important questions about its dynamical influence on the inner planetary system. The Kepler-444 system addresses many challenges to the formation and evolution models. It renders this system particularly interesting, and the delivery of the most precise orbital parameters and masses would definitely be a valuable information for future studies. 

Kepler-444\,A is a bright (V=8.86), K-type star of mass $m_{\star}$\,=\,0.758\,$m_{\odot}$ and radius $R_{\star}$\,=\,0.752\,$R_{\odot}$ \citep{Campante2015}. It displays significant pulsating signals, and thus constitutes a precious laboratory for the asteroseismic models. Using such models, \citet{Buldgen2019} confirm the old age of the system (11$\pm$0.8\,Gyr), in accordance with previous kinematic and Ti abundance studies which indicate that the system belongs to the galactic thick disk \citep{Campante2015}. 
The stellar companions B and C have estimated masses of 0.29\,m$_{\odot}$ and 0.25\,m$_{\odot}$ according to \citet{Dupuy2016} and using mass-magnitude relationships. Furthermore, they orbit each other very tightly, with an estimated semi-major axis of 0.3\,AU. The BC center of mass performs a full revolution in the reference frame of the primary in about 200 years, and on a highly eccentric orbit of $\sim$\,0.86, according to the mentioned study. These results imply that the M-dwarfs pair periodically passes within 5 AU of the primary. Therefore, significant gravitational perturbations on the planetary orbits potentially take place. Furthermore, if primordial, the presence of this highly eccentric stellar doublet would have truncated the protoplanetary disk in which the planets formed to an estimated radius of $\sim$\,2 AU \citep{Dupuy2016}, placing strong constraints on the planet formation processes. Several scenarios could explain the current architecture of the triple-star system. Either the observed configuration is primordial, or it was built later on under the effect of gravitational interactions. However, \citet{Dupuy2016} analysed the orbit of the stellar binary BC with respect to the primary A. They found a probability of 98$\%$ that this orbit lies in the same plane than the planetary orbits. Therefore, the authors conclude that the observed triple-star configuration is most likely primordial.  

The Kepler-444 sub-Venus planets would then have formed in a truncated disk of Kepler-444 A, given the high probability of primordial orbital configuration of the BC stellar companions around the primary. 
The formation process of the planets in this disk and their orbital evolution remain open questions. \citet{Papaloizou2016} investigated the smooth disk-driven migration of type I - given the small planetary masses - to explain the current system's architecture. The authors were able to reproduce the current architecture under this process. However, for it to be effective, planet $e$ has to be approximately three times more massive than planet $d$ in order to bring this pair close to mean-motion resonance (MMR) while keeping the other pairs far from such configuration. This condition is inconsistent with the TTV analyses of \citet{Mills2017}, that find planetary masses and densities of 0.036 and 0.034 $m_{\oplus}$ and 1.27 and 1.08 g/cm$^3$ for planets d and e, respectively. These authors suggest instead that either the migration was not smooth - e.g. because of the local disk properties or turbulences - or large perturbations arose in the system after the disk dispersal, significantly modifying the period ratios. Nevertheless, they note that given the low densities of planets $d$ and $e$, any collisional stripping process was very unlikely to take place as it would leave planets on higher densities. Finally, the planets may also have formed in situ. Having insights into the atmospheric composition of the planets would definitely help to refine and precise the formation model for this outstanding system. 

Given the small size of these planets and subsequently the shallowness of their transit signals, detecting features from the lower atmosphere via infrared transmission spectroscopy is out of reach with the current instruments. However, \citet{Bourrier2017} detected with the Hubble Space Telescope (HST) significant absorption of the stellar Lyman-alpha line at the times of transit of the outermost planets e and f. Two different scenarios could explain these features. 
These flux variations could come from strong and sudden activity variations of the star.
The second scenario proposes that these variations come from the absorption of light by neutral hydrogen in the extended upper atmosphere of these planets. \citet{Pezzotti2021} showed that the match between the stellar models and the asteroseismic observations is incompatible with an active star, 
confirming previous results from \citet{Bourrier2017} that Kepler-444 should be a quiet old star without any transient features. 
\citet{Pezzotti2021} thus favour the second scenario, in which neutral hydrogen would escape from the outer planets' atmospheres. 
A possible origin for this hydrogen is steam, photodissociated by stellar radiation in the upper planetary atmospheres, and sustained by water from the planet surface \citep{Jura2004}. Given the age of the system, large primordial water mantles would be required for the planets to still show signatures of escape today \citep{Bourrier2017}. This scenario is consistent with the low mean densities measured by \citet{Mills2017} for planets d and e. If confirmed, it would invalidate an in-situ formation scenario for the planetary system, given the rarity of water in the inner protoplanetary disk. 

There remains many open questions about Kepler-444. Follow-up formation studies will certainly benefit from the up-to-date orbital parameters of the system. Providing these revised parameters is the first goal of this work. 
Furthermore, questions of dynamical nature can also provide insights into the formation process. How does the compactness of the planetary orbits impact their stability? In other words, how close from orbital instability lies the planetary system? To which extent does the stellar binary companion perturbs the inner system? This paper attempts to answer these open questions. Sect. \ref{Sect:Without_Bin} presents an analysis on the planetary system, without the influence of the binary companion. We review the planetary parameters in light of orbital stability, and investigate MMR. We then review in Sect. \ref{Section:BinaryOrbit} the orbit of the stellar binary companion Kepler-444 BC, based on additional observations from Gaia and HIRES. We use those updated results in a dynamical model that includes the BC companion, and undertake a full revision of the planetary and BC companion parameters in Sect. \ref{Section:With_Bin}. Furthermore, using HST data, \citet{Bourrier2017} depicted a strong absorption of the stellar Lyman-alpha line uncorrelated with any of the planetary transits. Hence, the authors suggest that this feature may come from a yet-unknown sixth planet on a grazing orbit configuration, and with a large exosphere. We develop this further in Sect. \ref{Section:Planet6}, via investigations of the dynamical plausibility of a sixth planet scenario. Finally, we synthesize and discuss our results in Sect. \ref{Section:Conclusions}.

\section{Dynamical constraints on model 1: No binary companion BC}
\label{Sect:Without_Bin}

\subsection{Stability-driven refinement technique}
\label{SubSect:Stability-drivenTechnique} 
In \citet{Stalport2022}, we presented an approach to refine the planetary masses and orbital parameters in multi-planet systems based on stability arguments. In particular, this technique asserts the orbital stability based on heuristic arguments linking instability to dynamical chaos. So far, there does not exist a formal relationship between these two concepts. Yet, they are empirically linked as many studies emphasized \citep[e.g.][]{Chambers1996, Murray1997, Obertas2017, Rice2018, Hussain2020}. For the stability estimation, we use the Numerical Analysis of Fundamental Frequencies (NAFF) fast chaos indicator, together with a calibration procedure for the latter. The approach couples the efficient parameter space exploration with the fast stability estimation of NAFF, in order to perform importance sampling on the distributions of solutions for the planetary systems. 

The NAFF quantification of chaos is extensively described in \citet{Laskar1990, Laskar1993}, and synthesized in \citet{Stalport2022}. It exploits the results from secular non-chaotic dynamics that the semi-major axes of the planetary orbits stay constant on average. This is not true any more in the presence of chaos. The NAFF chaos indicator uses the results of numerical integrations, and in particular we use even time series of the planetary mean longitudes. In this work, we derive the planetary mean longitudes in Jacobi coordinates. From this time series, the frequency analysis technique is used to decompose the planetary motion in its constituent frequencies and estimate the mean-motion averaged over the timespan of the time series. The mean-motion is the frequency equivalent of the orbital period: $n=2\pi/P$. It thus benefits from a precise computation in the frequency domain, and is equivalent to a measurement of the semi-major axis given Kepler's third law. We split the total integration time in two halves, and employ frequency analysis over each part. The difference in the average mean-motion between the two halves of the integration directly informs about the level of chaos of the considered planetary orbit. Large differences are indicative of a significant drift in the mean-motions, thus revealing chaos. We define the NAFF of the system as the difference between the average mean-motions normalised over the initial mean-motion of the considered planetary orbit: 
\begin{equation} \label{eq:NAFF}
\mathrm{NAFF} ~ = ~ \max_{j} ~ \left[ log_{10} \dfrac{\mid n_{j,2} - n_{j,1} \mid}{n_{j,0}} \right] 
\end{equation}
where $n_{j,1}$ and $n_{j,2}$ are the averaged mean-motions over the first and second half of the integration respectively, for every planet $j$ in the system. $n_{j,0}$ is the initial Keplerian mean-motion of planet $j$. The maximal drift among all the planetary orbits is defines the chaos level of the whole system. The convergence of this chaos indicator is usually reached within $\sim$10$^4$ orbital revolutions of the outermost body. 

From a system's posterior distribution exploring the whole parameter space, we select a sample of system's configurations on which we quantify the chaos via the NAFF computation. These numerical integrations serve first at calibrating the NAFF indicator for the considered dynamical system, and then to apply importance sampling. The calibration procedure is based on the hypothesis that the real system is stable. When modelling the latter, it is common to obtain two dynamically distinct populations of systems: a weakly chaotic population, and a population harbouring stronger chaos. This is illustrated by displaying the NAFF distribution of the sample. In \citet{Stalport2022}, it was stressed that the strongly chaotic population is depleted in favour of the weak chaos when additional observations are added and a better modelling of the system is hence available. Therefore, the calibration strategy consists in defining the NAFF-stability threshold as the bottom of the population of stronger chaos in-between the two populations. This calibration strategy has the advantage of not necessitating additional numerical integrations in order to perform importance sampling. From this same distribution we then remove all the system's configurations with NAFF larger than the defined threshold. 

In this paper, when mentioning the stability-driven refinement technique we refer to the process described above: on a sample of the posterior distribution, we proceed to the NAFF computation and calibration, and then undertake importance sampling in order to derive a stable posterior distribution. Let us note that the presence of the nearby doublet of M stars on a very eccentric orbit around the primary limits us in the choice of the integrator, preventing any symplectic scheme. We opt for \texttt{IAS15} \citep{Rein2015}, a 15th order adaptive time-step integrator implemented in the \texttt{REBOUND} python package\footnote{\texttt{REBOUND} is a package to compute the dynamical evolution of N-body systems. It can be found at \url{https://rebound.readthedocs.io/en/latest/} and contains a large variety of numerical integrators.} \citep{Rein2012}. In order to properly compare the results with and without the stellar binary in our model, we use the same integrator throughout this work except if stated otherwise. 
A correction from general relativity was added in the numerical integrations following the formalism of \citet{Anderson1975}, and implemented in the python module \texttt{REBOUNDx} \citep{Tamayo2020}.

\subsection{Global analysis}
\label{SubSection:Without-Global}
We performed first a stability analysis of the Kepler-444 system without the stellar binary in our dynamical model. This study serves as a reference to later explore the net effect of accounting for the binary, but also, it provides insights on the impact that dynamical constraints have on this compact system without additional perturbation. 

To that aim, we downloaded the TTV posterior derived by \citet{Mills2017}, which contains the planet parameters as constrained by the TTV analysis, and extracted a sub-sample of 28\,000 system configurations. About 850 solutions are effectively mutually independent among this set, as the authors of the former study note. The reason for choosing a larger sample is to obtain statistically relevant results. 
Let us note that the eccentricity distributions are derived from the sum of two independent Gaussian distributions, for $\sqrt{e}\cos\omega$ and $\sqrt{e}\sin\omega$ respectively. 
The result favours low eccentricities, in accordance with other observed compact planetary systems and numerical simulations \citep{Fabrycky2014, Pu2015}. 
Also, the longitudes of the ascending nodes of all the planets are fixed to zero.

\begin{figure}
    \centering
\includegraphics[width=\columnwidth]{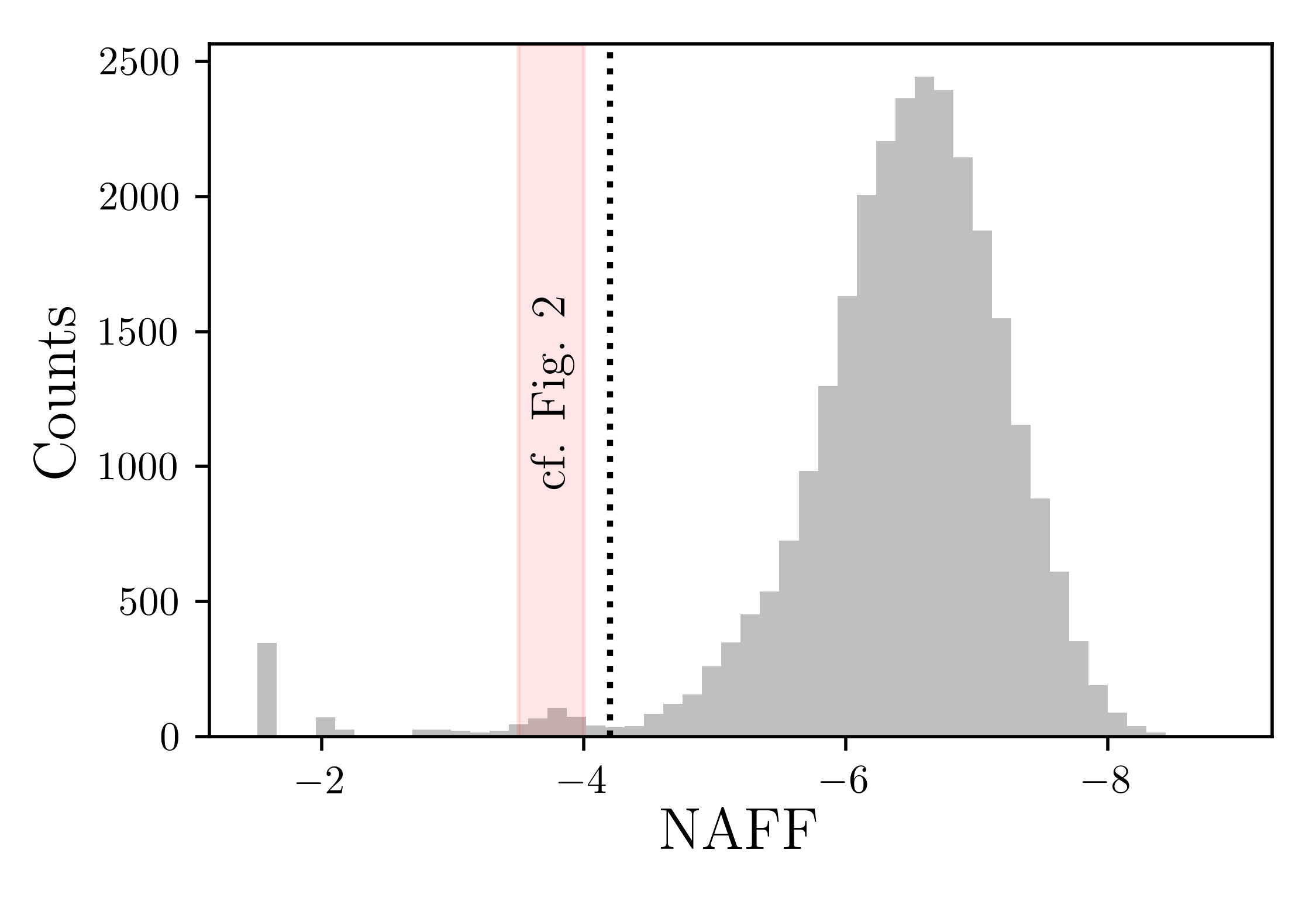}
    \caption{NAFF distribution of the Kepler-444 planetary system, based on the model posterior of \citet{Mills2017}. The vertical dotted line indicates the position of our threshold of orbital stability. Any system configuration with a NAFF chaos estimation smaller than this threshold (right-hand side) is flagged stable.}
    \label{fig:DistribNAFF_Kepler-444}
\end{figure}

Each of the 28\,000 solutions was numerically integrated with \texttt{IAS15} over 10 kyr - i.e. $\sim$ 375\,000 revolutions of the outer planet. 
We note a good precision of the integrations, with a relative total energy variation between $10^{-14}$ and $10^{-13}$. In every dynamical simulation, we record the planets' mean longitudes at regular time steps for a total of 20\,000 output times and use these time series to compute the NAFF chaos indicator on all the configurations that survived the entire integration (no close encounter, no escape). The criterion used for close encounter, i.e. for the minimal mutual distance acceptable between a planet pair, is the maximal initial Hill radius among the planets, which corresponds to the outermost one. Such a criterion prevents our integrator from employing excessively small time steps. Furthermore, once this distance is reached, it takes a small time before the collision occurs \citep{Rice2018}. Alternatively, we flag an escaped planet if its distance to the central star reaches five times the initial distance between the outermost planet and the star. In that case, we consider that the system architecture diverged too much from the initial configuration, and we stop the integration. 
The numerical simulations show a high level of stability. In Fig. \ref{fig:DistribNAFF_Kepler-444} we plot the histogram of the NAFF distribution that we derive. It is composed of 27\,890 values, which is the number of system configurations that survived the entire integration (no close encounter, no escape). Three families of systems are observed, at NAFF\,$\sim$\,-1.7, -3.8 and -6.5, corresponding to increasing regularity in the motions.
Following the calibration strategy described in \citet{Stalport2022}, we expect that if we had more observations that would further constrain the Kepler-444 planetary system, the peak in the distribution at NAFF\,$\sim$\,-1.7 would be further depleted in favour of the peak at NAFF\,$\sim$\,-6.5 in the new set of configurations. 

\begin{figure}
    \centering
\includegraphics[width=\columnwidth]{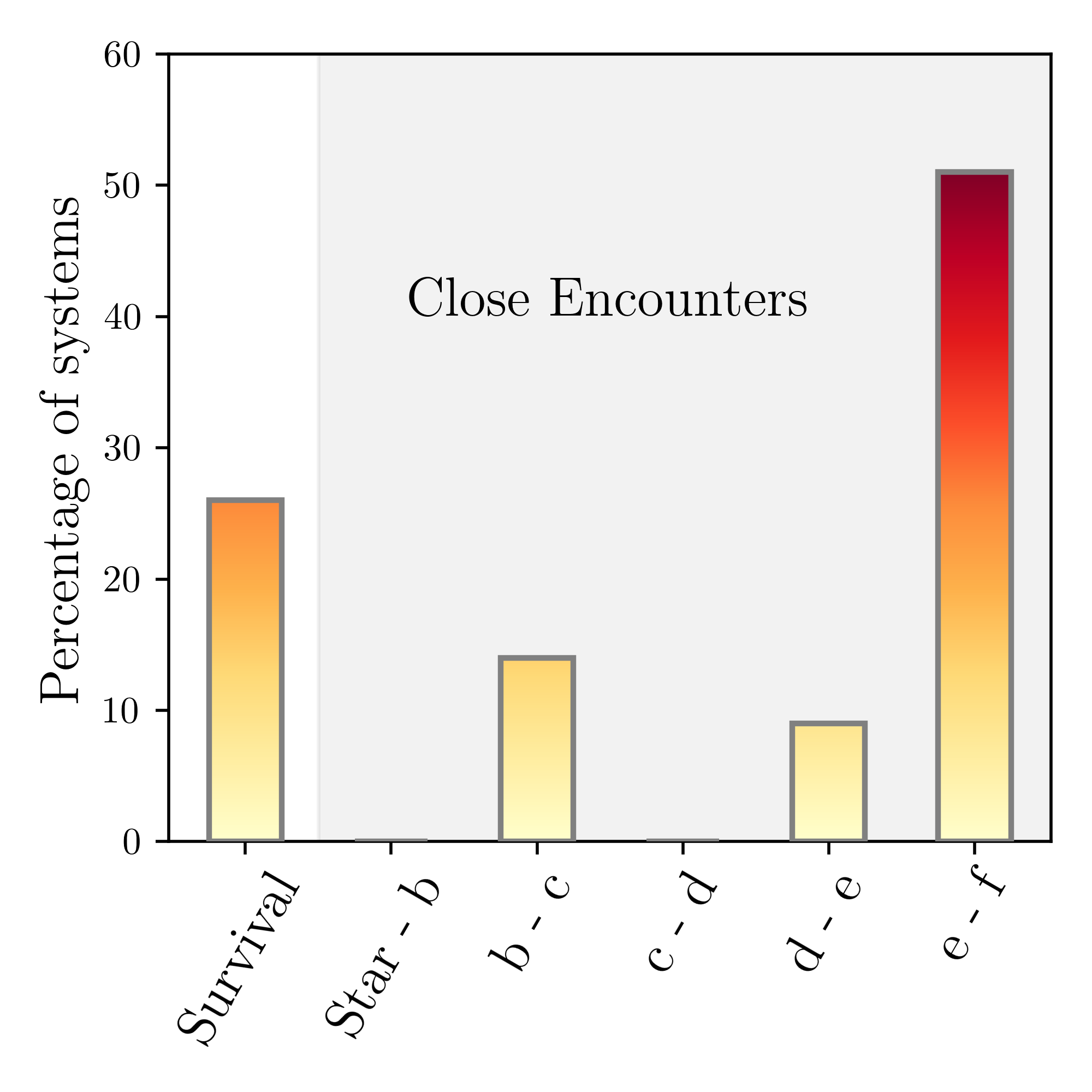}
    \caption{Fate of 100 system configurations that belong to the population of intermediate chaos peaking at NAFF$\sim$-3.5 (cf. red vertical band of Fig. \ref{fig:DistribNAFF_Kepler-444}). This bar chart distinguishes the systems that survived the 10Myr numerical integrations from those that encountered a close encounter, after which the simulation was stopped. Among the close encounters, we further differentiate between the bodies involved in the event.}
    \label{fig:Stat_LongInteg}
\end{figure}

The orbital stability of the systems population at NAFF $\sim$ -3.8 is unclear. To further investigate the dynamical behaviour of that population, we selected a hundred system configurations inside the red vertical band represented in Fig. \ref{fig:DistribNAFF_Kepler-444}. The latter spans the peak, and is hence representative of that intermediate population. Then, we computed the orbital evolution of each system over 10Myr, i.e. $\sim$375M of orbits of the outermost planet. For those long integrations, we employed the symplectic \texttt{WHFAST} integrator implemented in REBOUND \citep{Rein2015WHFAST}, together with a symplectic corrector of order 7 and a fixed time step of 1/100 of the innermost orbit. The integrations were stopped in case of close encounter or escape, using the same criteria as previously described. None of our integrations underwent a planet escape. Out of the 100 configurations, 26 made it to the end without instability. On the remaining 74 systems, there were 14 close encounters between planets b and c, 0 between planets c and d, 9 between planets d and e, and 51 between planets e and f. No close encounter between the central star and the innermost planet was recorded. Fig. \ref{fig:Stat_LongInteg} synthesizes those statistics. It is interesting to note that among the systems that suffer from an instability, most of them involve the outermost planet pair e-f. That indicates that this pair is the most likely to suffer from instability. 
Most of the systems had a close encounter within 10Myr, and a proportion of which did not may get unstable on longer time-scales. Therefore, we flag this intermediate population as unstable. 

Going back to Fig. \ref{fig:DistribNAFF_Kepler-444}, we set the NAFF-stability threshold at the bottom of the stable population -- i.e. the rightmost peak -- and opt for NAFF\,=\,-4.2. That means any system is flagged as unstable if the relative variations in the mean-motion of any planet is larger than about 0.006$\%$ over the two halves of the integration. Such an apparently small number is justified by the compactness of the planetary orbits. Out of the 28\,000 system configurations, 26\,988 survived the total integration time and were also flagged as NAFF-stable (NAFF < -4.2). Our NAFF stability criterion hence discarded an additional 902 system configurations, i.e. about nine times more than with the close-encounter criterion only. Overall, about 96$\%$ of the systems pass our stability criterion, which demonstrates that the posterior distribution obtained by \citet{Mills2017} is globally stable.

We compared their distribution with the new one, composed of NAFF-stable configurations only. As expected, they are no notable differences between the two given the overall stability of the original posterior. Notably, we note that the planetary masses do not get further refined by the orbital stability constraints. 
Additionally, despite the orbital compactness of the system, there is no clear impact of the stability constraint on the eccentricities. This is because the posterior distributions derived by \citet{Mills2017} are very much confined close to 0. Indeed, the 3$\sigma$ upper limits on the eccentricity of each planet, from innermost to outermost, are 0.070, 0.066, 0.069, 0.051, and 0.064. This is small enough to ensure orbital stability (as it will be further illustrated in Fig. \ref{fig:StMap}). As the posterior distributions are confined to low eccentricities, we do not expect the stability-driven approach to bring new constraints on the periastron directions, subsequently ill-defined. Indeed, we do not observe any constraints in the NAFF-stable distributions for the arguments of periastron. The mean longitudes are not further constrained by the dynamics either. This is no surprise since the Kepler transit observations provide already tight constraints on the orbital phases. Finally, the mutual inclinations between the different planetary orbits are not large enough to trigger orbital instability. Hence, we do not observe new constraints on the inclinations from the stability-driven approach.

\subsection{Resonant configuration}
\label{SubSection:MMR}
\begin{figure}
    \centering
\includegraphics[width=\columnwidth]{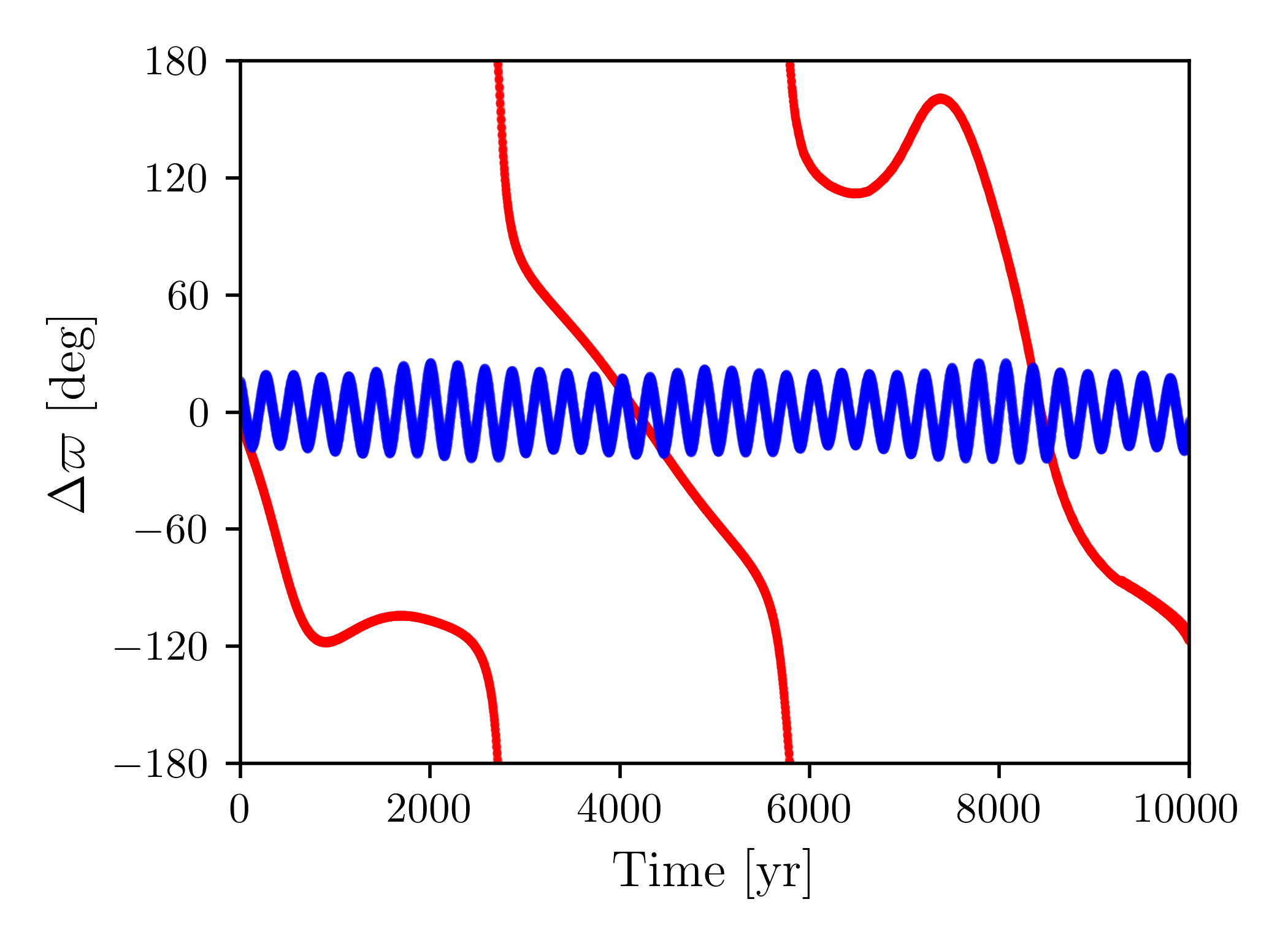}
    \caption{Temporal evolution of the difference in the longitudes of periastron $\varpi_1-\varpi_2$. This plot shows the evolution of two system configurations, which illustrates the diversity of dynamical behaviours -- libration (in blue) and circulation (in red).}
    \label{fig:Dw_NoBin}
\end{figure}

The planets pair d-e presents a period ratio very close to 5/4: $P_e/P_d$ = 1.251. The potential for this pair to lie in the 5:4 MMR may have an impact on the overall stability of the system. 
Our objective is double. First, we aim at unveiling the (non-)resonance nature of this d-e pair. Indeed, despite its proximity to the 5:4 MMR, the small planet-to-star mass ratios may induce too small resonance widths for the pair to be in resonance. Secondly, we aim at exploring the impact that the proximity to the 5:4 MMR has on the overall stability of the planetary system. 

Two resonant angles are associated with this first-order mean-motion resonance: 
\begin{equation} \label{eq:5:4MMR}
\begin{split}
\Theta_1 ~ & = ~ 5\lambda_e - 4\lambda_d - \varpi_d \\ 
\Theta_2 ~ & = ~ 5\lambda_e - 4\lambda_d - \varpi_e , 
\end{split}
\end{equation} 
where $\lambda$ is the mean longitude and $\varpi$ is the longitude of periastron. If the orbital periods are near commensurate -- i.e. if 5P$_d\sim$4P$_e$ -- the resonant angles vary slowly, and so is the case of their difference $\Theta_1-\Theta_2~=~\varpi_e-\varpi_d~\equiv~\Delta\varpi$. Among the system configurations that survived our 10kyr integrations, we found different dynamical regimes. Fig. \ref{fig:Dw_NoBin} illustrates that by showing two examples. In one case, the periastra of the d-e pair stay aligned with some oscillation. In the other case, the periastra are misaligned and their difference is not confined to any value, but instead, it circulates. The apparent libration of $\Delta\varpi$ can be misleading though, as it does not guarantee the resonance \citep[e.g.][]{Henrard1983}.  

\begin{figure} 
    \centering
\includegraphics[width=\columnwidth]{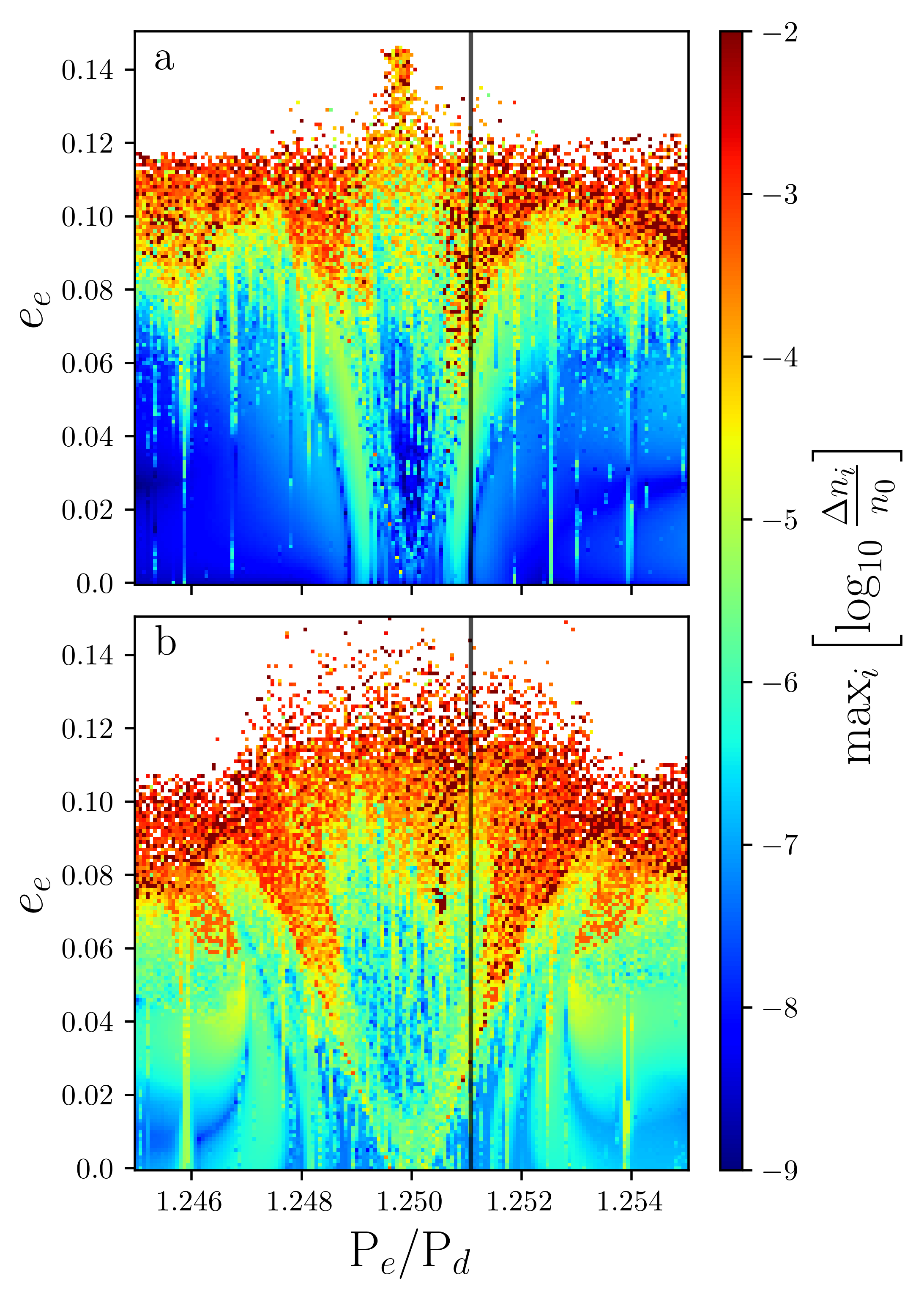}
    \caption{Chaoticity maps of Kepler-444 near the 5:4 MMR of the d-e planets pair. The topology of the parameter sub-space is revealed by a grid of 151x151 system's configurations and a color code given by the level of chaos in each configuration estimated via NAFF (cf. Sect. \ref{SubSect:Stability-drivenTechnique}). The redder areas correspond to strongly chaotic configurations, as opposed to the blue ones. Finally, the boxes are coloured in white if the system turned unstable before the end of the integration (either ejection or close-encounter). The vertical line indicates the position of the actual planets pair on this x-axis, as tightly constrained by the transits observations. The two grids differ from each other through the masses of planets d and e. Top (a): the planetary masses are fixed at the median of the TTV posterior obtained by \citet{Mills2017}. Bottom (b): the masses of planets d and e are increased to their 1$\sigma$ upper limits.}
    \label{fig:StMap}
\end{figure}

In order to investigate deeper the resonant behaviour of the d-e planet pair, we computed a chaoticity map exploring the influence of the period ratio $P_e/P_d$ on the x-axis and the eccentricity of planet e on the y-axis. 
Let us note that this choice of parameters for the map is somewhat arbitrary. However, in this sub-space we expect the first-order resonances between planets $d$ and $e$ to appear as clear V-shaped structures, rendering easier their visual identification. 
All the other orbital parameters of the planets were initially fixed from the median of the posteriors given in \citet{Mills2017}, except for the orbital inclinations that were all fixed at 90 deg. The masses of these two planets d and e were also fixed from that mentioned study. The masses of the other planets, unconstrained by TTVs, were estimated from the best radii estimations and with the Mass-Radius relation taken from \citet{Otegi2020}: $M = 0.9 R^{3.45}$ for the rocky exoplanets' population\footnote{Let us note that the Kepler-444 planets are unique by their very small sizes, all of them smaller than Venus. No M-R relationship for exoplanets has been validated so far in this size regime, and we thus make an extrapolation on the applicability of this relation.}. We initially varied only the eccentricity of planet e, $e_e$, in [0,0.15], and the period of that same planet, $P_e$, such that $P_e/P_d$ swipes the interval [1.245,1.255]. As such, we created a grid of 151x151 system's configurations, that we integrated using the same set-up as described previously. The NAFF results are used this time as a color code - not as a binary stability criterion. It confers a topology to the parameter space. The resulting map is presented in Fig. \ref{fig:StMap}a. 

The resonance appears clearly. It harbours this typical V-shape island of stability in the eccentricity - period sub-space, and is surrounded by layers of usually strong chaos around the separatrix that delimit the resonance. However, the separatrix does not harbour strong instability in this case, or it is too thin to appear clearly on this map - its average level of chaos is globally less strong than our stability threshold of NAFF\,=\,-4.2 set in Sect. \ref{SubSection:Without-Global}. It therefore seems that the proximity of the planets pair d-e to the resonance is not responsible for the overall instability of certain configurations of systems. Instead, the impact of large eccentricities plays a larger role in generating orbital instability. Furthermore, with the map displayed in Fig. \ref{fig:StMap}a, it is still unclear if the planets pair d-e actually lies inside or outside the topology of the 5:4 MMR. With the planetary parameters used to construct this map, the pair lies slightly outside of the resonance with reasonably low, non-zero eccentricities. The transits observations with Kepler allowed to strongly constrain the periods. For instance, the estimated period ratio of the d-e pair is represented by the vertical line. The uncertainty on the position of the latter along the x-axis is smaller than the width of a pixel. 
However, other planet parameters such as the masses or orbital eccentricities have an impact on the resonance shape and are less well constrained. As a result, modifying the values of those parameters may place the d-e pair inside or outside of the 5:4 MMR. Fig. \ref{fig:StMap}b illustrates the impact that a change in the estimated masses of planets d and e would have on the dynamical state of the system. In this new map, the masses of these two planets were increased to their 1$\sigma$ upper limit. This modification leads to an increase in the width of the MMR, so that the d-e planet pair could actually lie inside the resonance and with reasonably low orbital eccentricities. Therefore we cannot conclude on the exact dynamical state of that pair, since several of such states are compatible within the parameters uncertainties.  
Finally, we note that the eccentricity of planet $e$ is very unlikely to be larger than 0.1, given the high level of chaos of those configurations. This is in agreement with the global study that we performed in Sect. \ref{SubSection:Without-Global}, where the eccentricities never reached 0.1 explaining why the stability constraints did not have a significant impact. 

We also investigated the possibility for the system to be trapped in a three-planet MMR. Those resonances involving three bodies likely play a significant role during the planets migration in the protoplanetary disk \citep{Charalambous2018, Petit2021}. Additionally, they are crucial to understand the development of orbital instability in tightly packed systems \citep{Quillen2011, Petit2020}. Low-order three-planet MMR are thinner but also populate the phase space more densely than two-planet MMR. A triplet of planets is said in a three-body MMR if there exists integers $k_1$, $k_2$, and $k_3$ such that 
\begin{equation}
k_1n_1~+~k_2n_2~+~k_3n_3~=~0. 
\end{equation}
The resonance is further classified as order zero if $k_1+k_2+k_3=0$, and order 1 if $k_1+k_2+k_3=1$. Resonances of higher order are increasingly thinner in the phase space. Their influence is therefore smaller and their detection is practically more difficult. We searched for the closest three-planet MMR of order 0 and 1, based on the median planet parameters derived from the NAFF-stable distributions. We list the five closest MMR in Table \ref{tab:3-bodyMMR_V2}.

\begin{table}
\caption{List of the closest 3-planet MMR, which satisfy $k_1n_1~+~k_2n_2~+~k_3n_3~=~0$ ($k_1$, $k_2$, $k_3$ integers).}
\label{tab:3-bodyMMR_V2}
\centering
\begin{tabular}{@{}ccccc@{}}
\hline\hline
Planets triplet  & $k_1$    & $k_3$ & $k_1n_1~+~k_2n_2~+~k_3n_3$ & MMR order  \\ 
\hline 
b - c - d  & 1 & 1 & -0.00283856 & 0  \\  
d - e - f  & 1 & 6 & 0.00348929 & 1   \\ 
d - e - f  & 4 & 5 & -0.00416141 & 0  \\ 
c - d - e  & 1 & 7 & -0.00618178 & 1   \\ 
d - e - f  & 2 & 7 & 0.00956999 & 1   \\  
\hline
\end{tabular}
\tablefoot{This table focuses on zero-order 3-planet MMR ($k_1+k_2+k_3=0$) and first-order 3-planet MMR ($k_1+k_2+k_3=1$). We limit our search for combinations such that |$k_1$|+|$k_3$|<10.}
\end{table}

\begin{figure} 
    \centering
\includegraphics[width=\columnwidth]{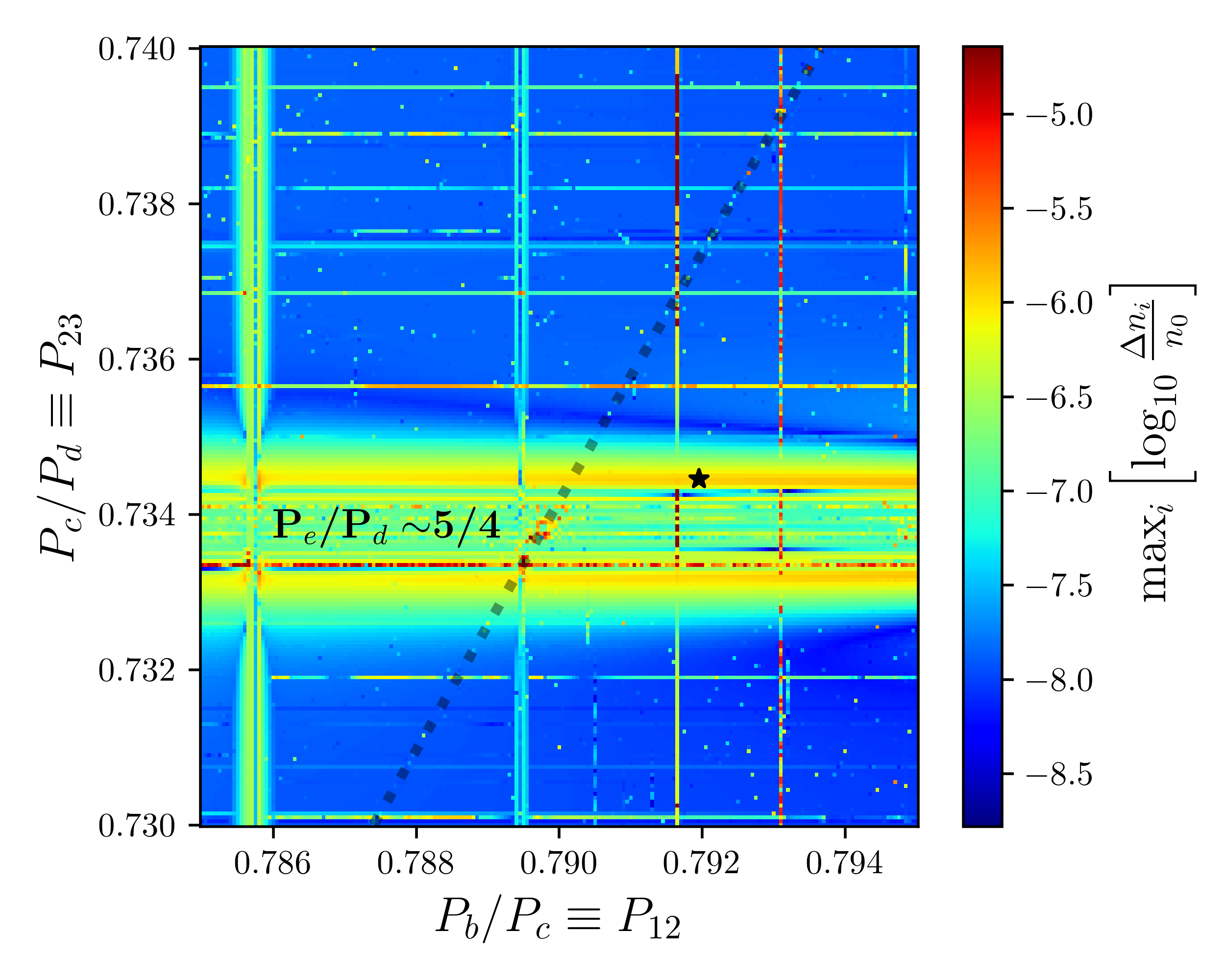}
    \caption{Chaoticity map of Kepler-444 in the space of the period ratios $P_{12}$ and $P_{23}$. The dotted line follows the expected location of the zeroth-order 3-body MMR $n_1 - 2n_2 + n_3 = 0$. The star depicts the position of the Kepler-444 b-c-d triplet in that space. The color scale is different than in Fig. \ref{fig:StMap}, in order to optimize the visibility of the different MMR.}  
    \label{fig:StMap_3MMR}
\end{figure}

The b-c-d triplet of planets is the closest to a low-order three-body MMR. Therefore, we focused on this triplet to study the potential 3-body resonant state in the planetary system. To do so, we studied the structure of the phase space in the plane of the period ratios $P_{12}\equiv P_b$/$P_c$ and $P_{23}\equiv P_c$/$P_d$. In that plane, 2-body MMR appear as vertical or horizontal bands, while 3-body MMR are oblique lines. Fig. \ref{fig:StMap_3MMR} presents the 201$\times$201 chaos map that we computed in this space, and the star marks the location of the Kepler-444 system. It reveals a web of two-planet MMR. Notably, the thick horizontal band is associated with the 5:4 MMR between planets d and e. Once again, we notice that the d-e planet pair lies at the outskirts of that resonance. The 3-body zeroth-order MMR that we identified at the first line of Table \ref{tab:3-bodyMMR_V2} is very thin. It is highlighted with a dotted line to guide the eye. This result shows that the b-c-d triplet is not in the considered 3-body MMR. Given that this triplet is the closest to a 3-body MMR, we conclude that none of the planetary triplets in the Kepler-444 system are in a 3-body MMR.

\section{Revision of Kepler-444 BC orbit} 
\label{Section:BinaryOrbit}

In this section, we assess the constraints on the orbital parameters of the Kepler-444 A-BC binary orbit. The BC component is currently near the apastron of a high eccentricity orbit, with \citet{Dupuy2016} finding e=0.864$\pm$0.023. That work used the relative astrometry of A-BC, the radial velocity trend of the A component, and the relative radial velocity (i.e. RV$_\textrm{BC}$-RV$_\textrm{A}$), and fixed the mass ratio (M$_\textrm{B}$+M$_\textrm{C}$)/M$_\textrm{A}$ to a value of 0.71.

There are three new sources of information that can potentially be used to further constrain the Kepler 444 A-BC orbit, namely:
\begin{enumerate}
\item Additional radial velocities from \citet{Butler2017}. While these significantly increase the density of RV points between 2012 and 2014, they do not extend the RV baseline relative to \citet{Dupuy2016}.
\item The astrometric acceleration of the host star, as derived from a comparison between the Hipparcos and Gaia catalog measurements of the stellar proper motion. Together, these two catalogs probe the astrometric acceleration of the star over a ~25yr baseline, which can be significant in constraining long-period orbits \citep[see e.g.][]{SnellenBrown2018,dupuy2019}.
\item The proper motion of the Kepler-444 BC companion, which is resolved from Kepler-444 A in the Gaia catalog \citep{gaia_mission, gaia_edr3}. We find, however, that this proper motion is not sufficiently reliable to use in the orbital fit, as discussed in Sect. \ref{subsection:bc_cpm} below.

\end{enumerate}

\subsection{Orbital fit using HGCA and additional RV data}
 
We fit a new orbit using only sources (1) and (2) of additional data, namely the new radial velocities and the astrometric acceleration of the host star; the companion proper motion is discussed in Sect. \ref{subsection:bc_cpm}. The full set of data used for the orbital fit is as follows: Keck/HIRES radial velocity values from \citet{Sozzetti2009}, \citet{Dupuy2016} and \citet{Butler2017}, Keck/NIRC2 relative astrometry and Keck/HIRES relative radial velocity measurements from \citet{Dupuy2016}, and absolute astrometry from the Hipparcos and Gaia measurements of Kepler-444 A. For the absolute astrometry, we used the EDR3 version of Hipparcos-Gaia Catalog of Accelerations (HGCA; \citealt{Brandt2021a}), which gives locally calibrated positional and proper motion differences between the two epochs.

To fit the orbit, we used the \texttt{orvara} orbit fitting tool \citep{Brandt2021b}, an efficient orbit fitter designed to simultaneously fit host star radial velocities, relative astrometry and absolute astrometry. We used the quoted measurements and errors from the works listed above, and an additional jitter term that was applied to all three RV instruments. We modified a branch of \texttt{orvara} such that it reads in a file listing relative RV measurements (i.e. RV$_\textrm{BC}$-RV$_\textrm{A}$), and calculates relative RV value(s) for each orbit model. The difference between the measured and calculated relative RV is incorporated into the \texttt{orvara} log-likelihood function, which is otherwise unchanged. This modified branch of the \texttt{orvara} code is available on github\footnote{https://github.com/ecmatthews/orvara}.

Unlike \citet{Dupuy2016}, we did not fix the mass ratio of A-BC. Instead, we used the asteroseismologically determined mass of 0.76$\pm$0.043M$_\odot$ for Kepler-444\,A (see \citealt{Campante2015}), and a broad, uninformative prior for Kepler-444\,BC, since the absolute astrometric information from the HGCA allows the mass of the secondary to be determined dynamically. We used the default priors as listed in \citet{Brandt2021b} for all other parameters. We ran 100 MCMC walkers for 50,000 steps, and discarded the first 10,000 steps as burn-in. Among all the parameters, the largest autocorrelation time that we obtained is 2,487.  

\begin{figure*}
    \centering
    \includegraphics[height=0.4\textwidth, trim=0cm 0cm 2.6cm 0cm, clip]{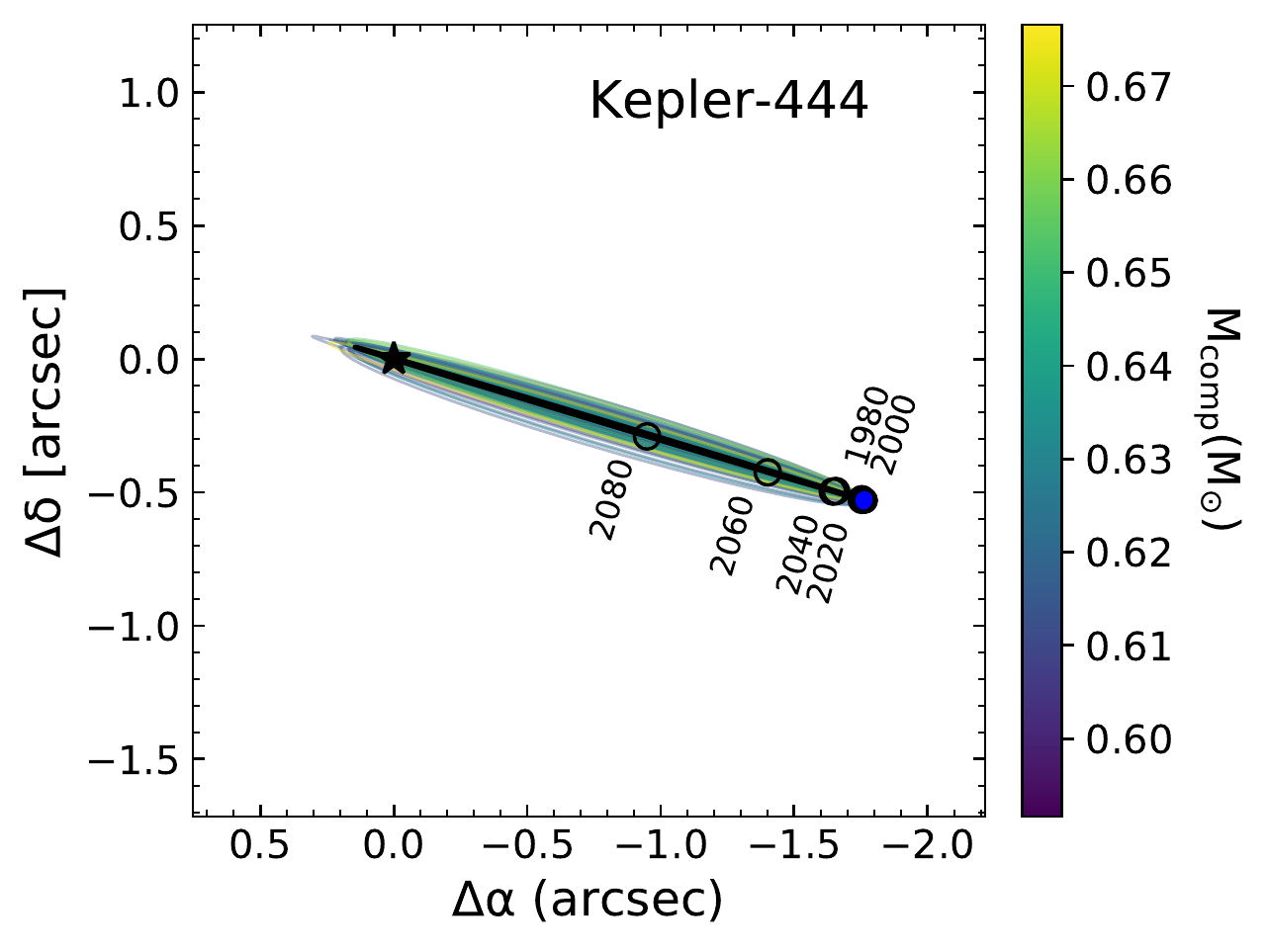}
    \includegraphics[height=0.4\textwidth]{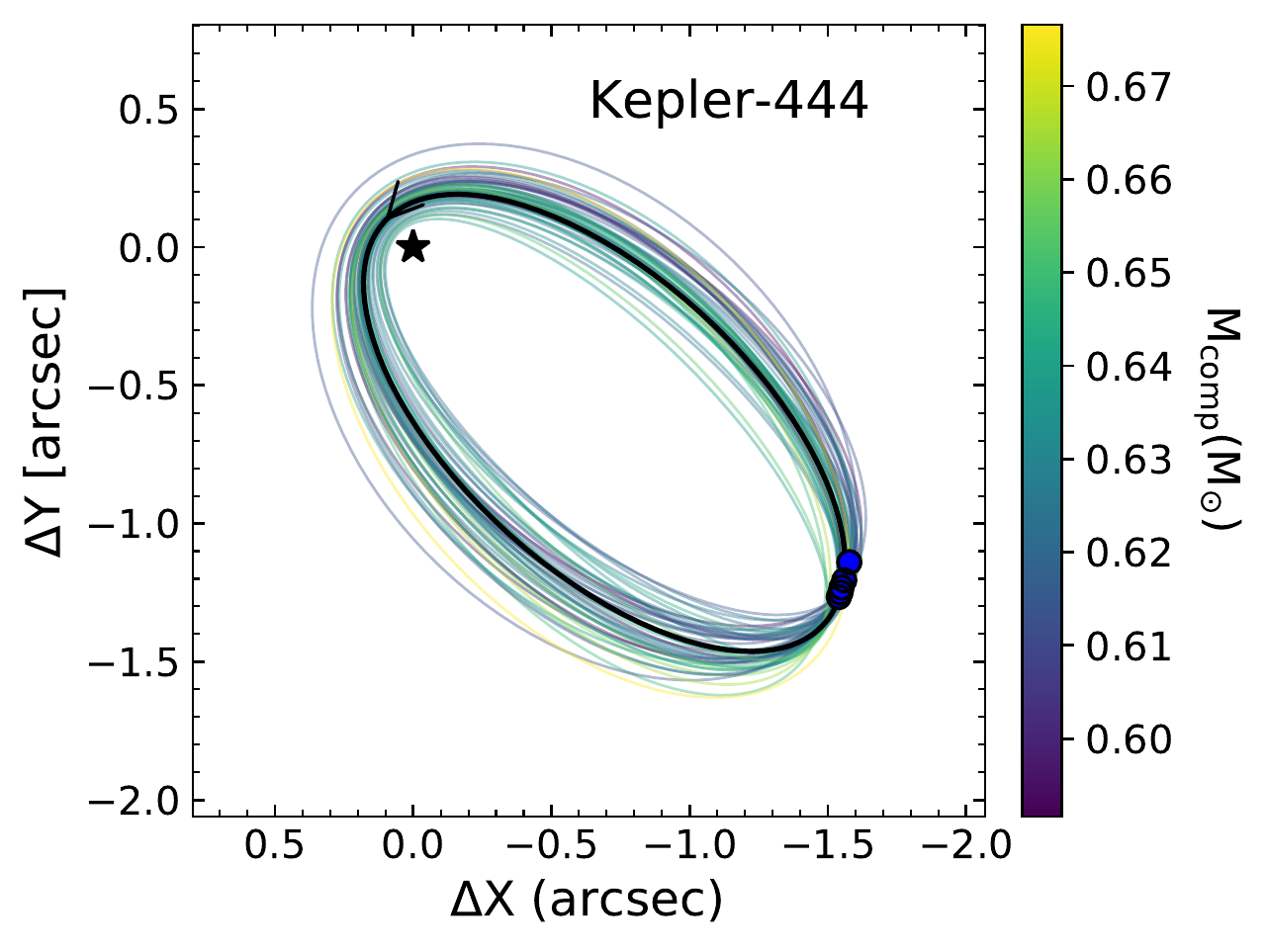}
    \caption{\textit{Left:} Best fit orbital solution (thick line) and 50 draws from the posterior distribution (thin, colored lines) for the Kepler-444 A-BC system, presented in the plane of the sky. Blue points indicate the observed relative astrometry of Kepler-444 BC, while open circles indicate predicted on-sky locations of the companion between 1980 and 2080. \textit{Right:} The same orbital solutions, presented in a top-down view. The observed astrometry is also deprojected into this view, using the inclination of the best-fit orbit. An arrow indicates the direction of this orbit. The colorbar indicates the total mass of the secondary component (i.e. M$_\textrm{B}$+M$_\textrm{C}$), and applies to both panels.}
    \label{fig:Kepler444orbits}
\end{figure*}

The updated orbital parameters are given in Table \ref{tab:binaryorbit}, and the orbit is plotted in Figure \ref{fig:Kepler444orbits}. We derive an eccentricity of ${0.865}_{-0.034}^{+0.031}$ and an inclination of ${90.6}_{-3.6}^{+3.7\circ}$: both of these values are in close agreement with the orbit presented in \citet{Dupuy2016}, confirming that the binary is indeed near apastron of an edge-on, eccentric orbit, and is compatible with being aligned with the primary.

Our approach yields a dynamically determined mass for the secondary of ${0.633}\pm{0.018}$M$_\odot$. This corresponds to a mass ratio of ${0.84}_{-0.05}^{+0.06}$, which is slightly higher than the fixed ratio of 0.71 used by \citet{Dupuy2016}, but remains consistent with photometry of the BC pair. We also find a marginally higher semi-major axis of ${39.5}_{-0.8}^{+1.0}$AU, and correspondingly a slightly longer period of ${211}_{-8}^{+9}$yr. The semi-major axis and period values are consistent to 2.3$\sigma$ and 1.1$\sigma$ respectively with the values given in \citet{Dupuy2016}.

\begin{table}
\centering
\caption{Revised orbital parameters for Kepler-444 A-BC.}
\label{tab:binaryorbit}
\footnotesize
\setlength{\extrarowheight}{4pt}
\begin{tabular}{ll}
\toprule
\textbf{Parameter [Unit]}  & \textbf{Value} \\ 
\midrule
Semi-major axis [AU] & ${39.5}^{+1.0}_{-0.8}$ \\
Eccentricity         & ${0.865}_{-0.034}^{+0.031}$ \\
Inclination [deg]    &    ${90.6}_{-3.6}^{+3.7}$ \\
PA of the ascending node [deg] & ${73.0}_{-1.4}^{+1.4}$ \\
Argument of periastron [deg]  & ${334.3}_{-1.8}^{+1.7}$ \\
Period [yrs] & ${211}_{-8}^{+9}$ \\
Time of periastron [JD] & ${2488532}_{-708}^{+785}$ \\

Closest approach at periastron a(1-e) [AU] & ${5.3}_{-1.3}^{+1.5}$ \\

Msec [M$_\odot$]    &     ${0.633\pm0.018}$ \\
mass ratio, (M$_\textrm{B}$+M$_\textrm{C}$)/M$_\textrm{A}$  &       ${0.84}_{-0.05}^{+0.06}$ \\
\bottomrule \bottomrule
\end{tabular}
\end{table}

\subsection{Kepler-444 BC proper motion}
\label{subsection:bc_cpm}

Kepler-444 BC is resolved from Kepler-444 A in the Gaia catalog, meaning that there is a direct measurement of its proper motion. However, there is reason to be cautious of the Kepler-444 BC proper motion measurement, since the Kepler-444\,BC binary is not only close ($<$2'') to its much brighter host star, but is also an unresolved binary. Indeed, the Gaia parallax of BC is inconsistent with the parallax of the A component at the 5$\sigma$ level. The quoted parallax values for the two components indicate that BC is radially separated from A by 0.33$\pm$0.07pc, or 68,000$\pm$15,000au. This is clearly inconsistent with the orbit derived above, which has a semi-major axis of just $\sim$40au, and suggests an issue with the astrometric solution for BC. Further, Kepler-444 BC has a remarkably high \texttt{ipd\_frac\_multi\_peak} value of 31 in the Gaia catalog. This value represents the percentage of CCD transits where additional images were seen \citep[see][]{Fabricius2021}, suggesting that either the BC pair is resolved in a subset of images, or a faint background star is confusing the images and biasing the astrometric solution. Together, these lines of evidence clearly indicate a system with an unreliable astrometric solution, and we therefore do not use the direct proper motion measurements of Kepler-444 BC when fitting its orbit.

This is in agreement with \citet{Pearce2020}, who did attempt to fit the orbit of Kepler-444 A-BC using only the Gaia DR2 direct measurements of Kepler-444 BC, and found that the astrometry of the companion was not sufficient to constrain its orbit. Their orbital fit for this companion is highly inconsistent with that of \citet{Dupuy2016}, and they infer that the astrometric solution is biased, since Kepler-444 BC is an unresolved binary.

\section{Dynamical constraints on model 2: With the binary companion BC} 
\label{Section:With_Bin}
\begin{figure*}
    \centering
\includegraphics[width=\textwidth]{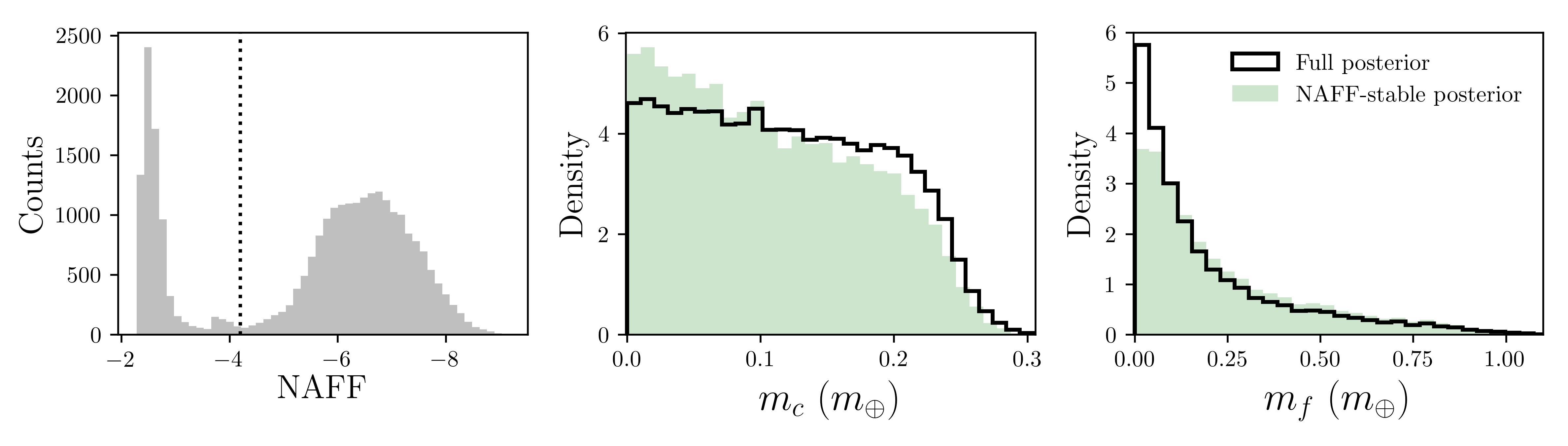}
    \caption{\textit{Left} -- NAFF distribution of the Kepler-444 system from the model posterior that includes the binary companion BC. The vertical dotted line represents our stability threshold. Any system configuration with a NAFF chaos estimation smaller than -4.2 (right-hand side) is flagged stable. \textit{Middle} -- Mass distributions of planet c, both presenting the full posterior distribution (black line) and the distribution of NAFF-stable configurations only (plain green).  \textit{Right} -- Same for planet f.}
    \label{fig:DistribNAFF-and-Mass_Kepler-444_WithBin}
\end{figure*}

We performed a global dynamical analysis with the stellar binary included in our model. For the sake of simplicity, we modelled the couple BC of M-type stars with a single body of mass equal to the sum of the masses of each component. The momentum induced by the separate components on the inner system is indeed expected to have very limited impact on the dynamics. This was further confirmed by \citet{Dupuy2016}, section 4.2, via numerical simulations. Furthermore, we apply a GR correction on the gravitational potential of the central star only, and the same formalism as the one used in Sect. \ref{Sect:Without_Bin} is employed. The stellar binary companion is indeed too far from the inner system for GR effects to play a significant role. 
The orbit of the binary companion was derived from our fit, described in Sect. \ref{Section:BinaryOrbit}. 
We built a global distribution of solutions combining the TTV posterior of \citet{Mills2017} for the planets with our posterior of orbit configurations for the stellar binary. Therefore, to each of the 28\,000 planetary system configurations that we extracted in Sect. \ref{Sect:Without_Bin}, we added a different solution for the orbit of the stellar binary. 
Every unique system configuration of the resulting global distribution was integrated over 100 kyr with \texttt{IAS15}. Then, we computed the NAFF chaos indicator on the configurations that survived the entire integration based on 20\,000 even records of the planetary mean longitudes. 
Again, most of the system configurations survive the entire numerical experiment -- out of the 28\,000 systems, 27\,311 make it to the end (no close-encounter, no escape). With a difference however, the NAFF distribution reveals a more contrasted picture. The left plot of Fig. \ref{fig:DistribNAFF-and-Mass_Kepler-444_WithBin} presents that distribution. The peak of the strongly chaotic population appears dramatically more significant compared to the model without binary. Therefore, the inclusion of the stellar binary companion BC in the model develops significant chaos in the inner planetary system.

From that NAFF distribution, we notice that the stability threshold is once again conveniently placed at NAFF=-4.2, and we keep only the systems for which their NAFF chaos estimate is smaller than this value. Removing the unstable systems, 19\,660 configurations pass the stability criterion. 
This sample builds up the stable posterior. 
This new distribution brings additional constraints, unseen without the stability analysis and absent from the dynamical model without stellar binary. As such, the masses of planets c and f encounter notable changes. The middle and right plots of Fig. \ref{fig:DistribNAFF-and-Mass_Kepler-444_WithBin} present their distributions before (black line histogram) and after (plain green) the sorting in NAFF. The stability constraint favours the lower end of the distribution for planet c, while the exact opposite is observed for planet f. Indeed, the gravitational perturbation that the latter encounters from the binary companion increases if the mass ratio decreases. As such, the smaller the planet mass with respect to the stellar binary, the stronger the perturbation and the more chaotic is the system. In the case of planet c, the dominant effect is different. The planet is embedded in the planetary system, i.e. it is surrounded by planetary companions. Therefore, the effect of the stellar binary is indirect. Instead, an increase in the mass of planet c directly results in stronger gravitational interactions with its planetary neighbours. This is particularly prominent in tightly packed configurations, hence further developing chaos. As such, the lowest values for the mass of planet c are the most likely to provide stable system configurations.  

Again, we note that the masses of close-to-MMR planets d and e are not further refined via the stability constraint. The best estimations remain those obtained via TTV analyses \citep{Mills2017}. 
This is in accordance with the dynamical studies of \citet{Tamayo2021} on Kepler-23, which showed the better ability for TTVs than stability constraints in refining the planetary masses. The other planetary dynamical parameters do not present significant updates with the stability constraint. We synthesize the results and provide a review of the planetary parameters in Table \ref{tab:Solution}. 

\begin{figure*}
    \centering
\includegraphics[scale=0.9]{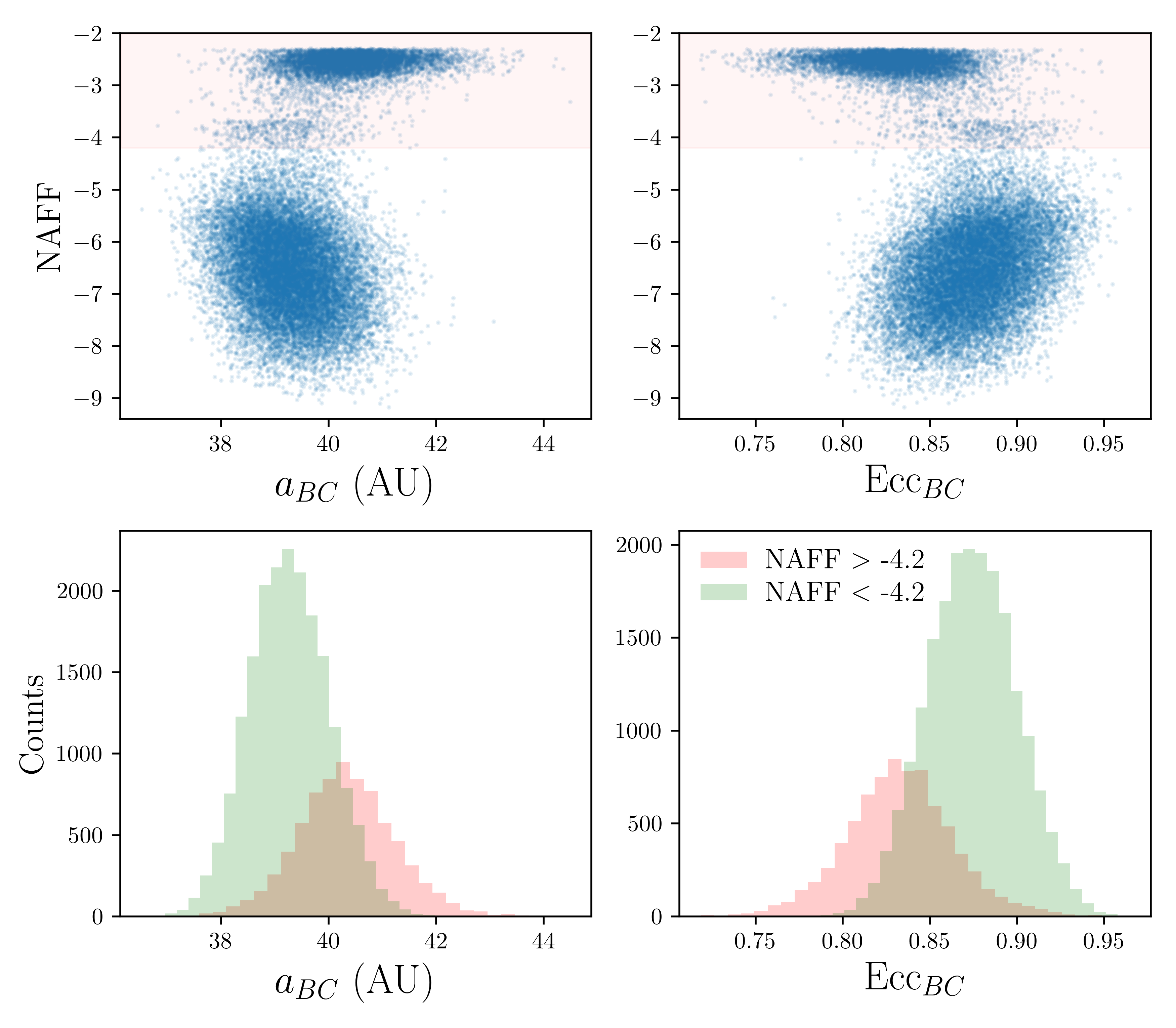}
    \caption{\textit{Top left} -- NAFF chaos indicator with respect to the semi-major axis of the stellar binary companion $a_{BC}$. The red zone depicts the area where the system configurations were flagged unstable. \textit{Bottom left} -- Distributions of $a_{BC}$ for both the stable (in green) and the unstable (in red) populations of systems. These two populations are delimited by the threshold value NAFF=-4.2. \textit{Top and bottom right} -- Same for the orbital eccentricity of the stellar binary companion Ecc$_{BC}$.}
    \label{fig:BCdistriutions_a-e}
\end{figure*}

As was noted, the stellar binary companion Kepler-444 BC is responsible for increasing chaos in the planetary system. Therefore, we explored potential correlations between chaos in the planetary system and the exact orbit of the stellar binary. We find a dependency of both the semi-major axis and orbital eccentricity on the chaos. This is illustrated in Fig. \ref{fig:BCdistriutions_a-e}. The top left plot presents all the system configurations that survived the numerical simulation, and represented in terms of the NAFF chaos against the semi-major axis of the stellar binary $a_{BC}$. The red area identifies the region of that space that is flagged unstable with our NAFF stability criterion. The stable configurations form an island, which further constrains the semi-major axis of the binary companion. The histogram on the bottom left further stresses this observation. It presents the $a_{BC}$ distribution for the stable (in green) and the unstable (in red) sub-populations of systems. These two populations significantly differ from each other. The stable distribution favours small $a_{BC}$ and is thinner than the unstable one. That latter effect is equivalent to a decrease in the orbit uncertainties. The same behaviour is observed for the orbital eccentricity of the stellar binary companion Ecc$_{BC}$, with the difference that the stable distribution favours large values. This anti-correlation finds a natural explanation in the observations. The projected distance between A and BC is well-constrained. The stellar companion currently is located around its apastron, at which the distance from the central star is given by $a(1+e)$. Therefore, measuring the distance between A and BC induces a degeneracy between the eccentricity and the semi-major axis of the BC companion, justifying the mirror effects between the two plots at the top of Fig. \ref{fig:BCdistriutions_a-e}. 

As a consequence of these opposite effects in the distributions of $a_{BC}$ and Ecc$_{BC}$, the stellar binary companion BC passes even closer to the central star at periastron than what was originally estimated. Taking the median values of both distributions, we find that the periastron distance -- i.e. the distance of closest approach to the central star -- of the stellar binary is 4.94 AU, which is within 5 AU. This result appeals for a careful study of the interactions between BC and the protoplanetary disk in which the planets formed, with the potential to refine the truncation radius and further constrain the expected planet compositions. This work is beyond the scope of the present study. 
We updated all the dynamical parameters of the binary companion BC, and present them below the planet parameters in Table \ref{tab:Solution}.

As a final note, we stress that about 70$\%$ of the system configurations pass our NAFF stability criterion. This percentage is large, indicative of a limited influence from the stellar companion on the inner planetary system. The planetary system is sufficiently embedded in the gravitational potential well of the primary star Kepler-444 A to avoid being decimated by the gravitational perturbation of the BC companion. This result suggests that the perturbative power of such companions on close-by eccentric orbits is limited, and we may expect to discover various planetary orbital architectures in multiple-star systems. This proposition has to be contrasted though, as we did not perform a detailed study of the dynamics on long time-scales. Secular effects may play a significant role, and we briefly discuss this issue in Sect. \ref{Section:Conclusions}.

\section{Dynamical insights for a sixth planet}
\label{Section:Planet6}
During the observation campaign of Kepler-444\,A with the Hubble Space Telescope, \citet{Bourrier2017} identified $\sim$\,20$\%$ flux decreases 
in the stellar Ly-alpha line associated with the transits of planets $e$ and $f$. In addition, they noticed an unexpected event: a strong $\sim$\,40$\%$ flux decrease in the same spectral range, but at a time uncorrelated with any transit of the known planets. To explain this observation, they propose the existence of a sixth planet on a grazing configuration and with an extended, giant exosphere comparable in size to the one around GJ\,436 b \citep{Ehrenreich2015}. With such a scenario, the authors were able to reproduce the observations. If the orbit of this hypothetical planet was co-planar with the average orbital plane of the other planets, its transit limit would lie at $\sim$\,19 days of orbital period ($\sim$\,0.13 AU). 

\begin{figure} 
    \centering
\includegraphics[width=\columnwidth]{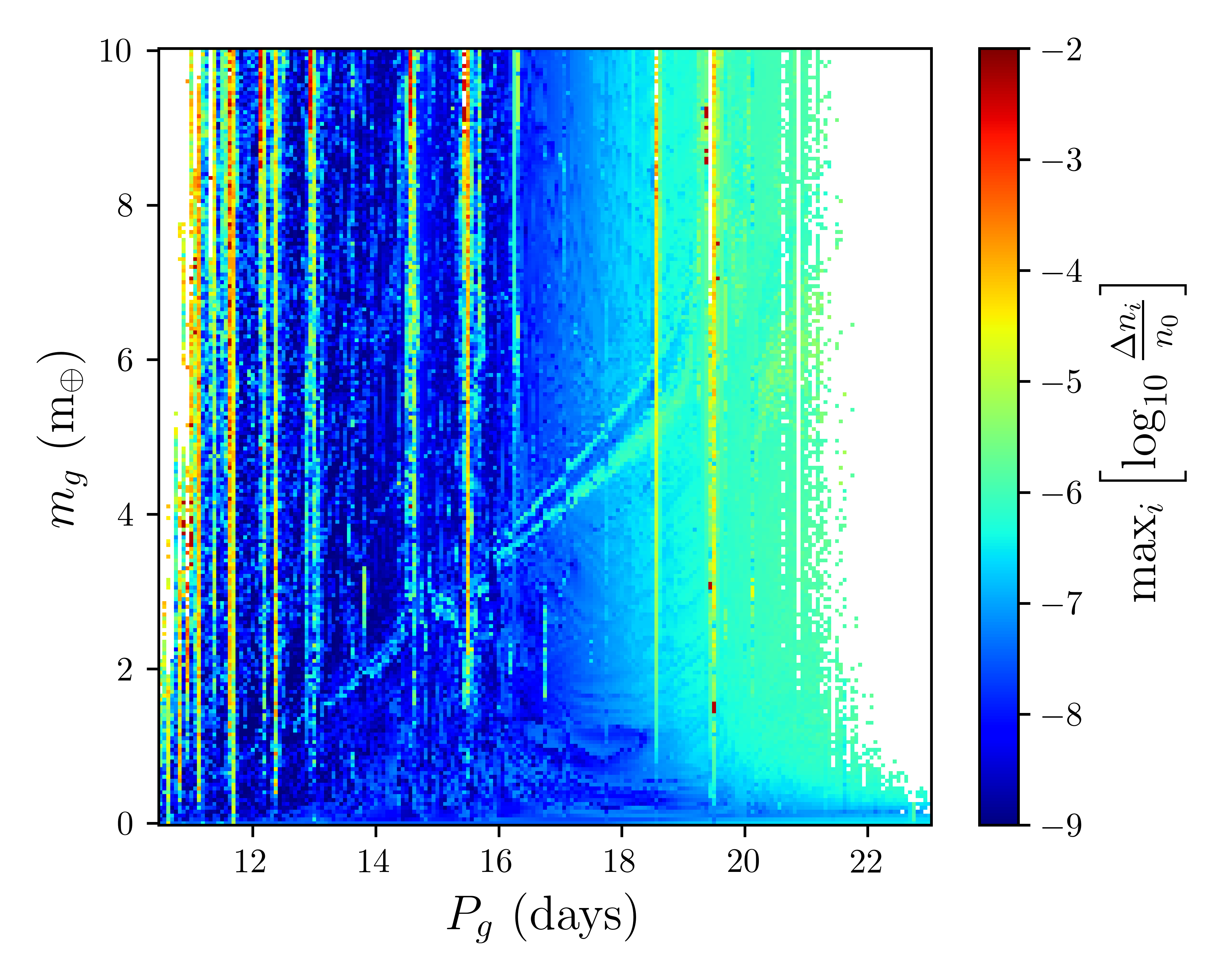}
    \caption{201$\times$201 chaoticity map of Kepler-444 using a model composed of six planets and the stellar binary companion (considered as a single body) orbiting the primary star. This map explores the plausibility for the existence of a putative sixth planet in the system, via the exploration of overall system chaoticity in the 2D space $m_g$ vs $P_g$. The level of chaos is provided by the NAFF indicator, and serves as the color code. White boxes indicate that the corresponding system configurations did not survive the entire integration (close encounter or ejection).}
    \label{fig:StMap_6thPlanet}
\end{figure}

In this section, we explore the dynamical plausibility for a sixth outer planet in the Kepler-444 system. We do not aim at carrying an exhaustive six-planet dynamical model, but seek at assessing the consistency of a putative planet $g$ on the orbital stability of the whole system. 
As such, we introduced a supplementary planet in the full dynamical model -- consisting of the primary star, the planets and the stellar binary BC -- with null eccentricity and in a edge-on orbit ($i$=90 deg). The other five planets were inferred null orbital eccentricity, while their orbital inclinations, transit times and orbital periods were taken from the results presented in Table \ref{tab:Solution}. The masses of these five known planets were inferred from the mass-radius relationship of \citet{Otegi2020}, in the same way as for the computation of the map presented in Fig. \ref{fig:StMap}. The BC parameters were taken from Table \ref{tab:Solution}. 
The orbital phase of the putative sixth planet in our simulations matches the observation of \citet{Bourrier2017}. 
The authors retrieved indeed the expected inferior conjunction time of the hypothetical planet based on the comparison between their exospheric model and the flux-decrease observations of the Ly-$\alpha$ line. They also note that around a solar-like star, the limit above which the atmosphere of a giant planet would be stable is 0.15 AU \citep{Koskinen2007}. That gives a crude estimate for the maximal semi-major axis of that hypothetical sixth planet, in order for the latter to harbour a large exosphere. That translates into an orbital period of $\sim$ 24.4 days. 
We computed a chaos map exploring the sub-space $m_g$ versus $P_g$, where the mass of the supplementary planet was varied between 0.001 and 10.001 Earth masses, while its orbital period was explored in the range [10.5,23.0] days. 
This was done via a grid of 201$\times$201 system's configurations, each of them integrated over 100 kyr. The NAFF indicator was computed at the end of the integrations, and served as the color code in order to depict the level of chaos in the system's configurations. The result is presented in Fig. \ref{fig:StMap_6thPlanet}. 

First, this map shows that there is room for a sixth outermost planet without destabilizing the whole system. The range of periods allowed is roughly comprised between 12 and 20 days. Below 12 days, the compactness of the f-g planet pair results in close encounters between the two bodies. Above 20 days, the stellar binary companion BC periodically exchanges significant momentum with the outermost planet during the passage at periastron. Consequently, planet g undergoes regular upward pushes of its orbital eccentricity, leading ultimately to the orbit crossing with planet f and a close encounter. In-between these limits, an additional sixth planet could harbour most of the tested orbital periods. 
These results thus validate the dynamical plausibility of the hypothetical sixth planet introduced in \citet{Bourrier2017}, and precise its range of possible orbital periods. In particular, a period window between $\sim$ 16 and 18 days shows low levels of chaos and corresponds to a planet g on a grazing orbit -- supposing its orbit to be co-planar with the average orbital plane of the other planets -- that would most likely have been missed in the Kepler observations. Furthermore, the mass of this planet has a very limited impact on the orbital stability. This leaves the possibility for a wide diversity of planets.  

We stress that these simulations investigate the short-term chaos in the system. Exploring the secular dynamics would be necessary to further study the orbital stability and refine the parameters of that potential sixth outer planet. This is beyond the scope of the present study.

\section{Conclusions} 
\label{Section:Conclusions}
The 11Gyr old triple-star system Kepler-444 constitutes a particularly challenging case study for the planet formation and evolution models. 
The binary Kepler-444 BC highly eccentric orbit around Kepler-444 A, and the tightly packed succession of sub-Venus planets orbiting the primary, raise many questions in terms of dynamics. How close to global instability is this compact architecture? How much does the binary companion perturb the inner system? These questions are of importance in understanding the system evolution from the formation stage until now. 
In this work, we reviewed the architecture of Kepler-444 in light of orbital dynamics, and tackled the questions above. 

We first investigated the dynamics of the inner system of Kepler-444, composed of the five planets and the primary star only. We inspected the orbital stability of the planets posterior distribution derived by \citet{Mills2017}, via the strategy introduced in \citet{Stalport2022}. We note a strong agreement with stability. Furthermore, we explored potential resonances in the planetary system. Notably, the planets pair d-e presents a period ratio close to 5/4. We revealed the 5:4 MMR in the phase space, but could not conclude about the pair being inside or outside that resonance. That strongly depends on the values used for the planet parameters. Additionally, we also investigated the three-planet MMR of order zero and one. None of the planet triplets lies in such resonances. The formation process of compact systems is expected to place the planets in a resonant configuration, via the disk-planet tidal interactions. This is not what is observed, which is consistent with the general results from the Kepler mission \citep{Lissauer2011,Fabrycky2014}. A potential explanation is the role of tidal dissipation with the star, which moves away the planets from the resonant configuration \citep{Delisle2014, Millholland2019}. Investigations of the potential impact of tides on the system dynamics would provide better insights on today architecture. 

As a second step, we analysed new observations in order to update the orbit of the stellar binary companion Kepler-444 BC, in prevision of its inclusion in our dynamical model. We used additional HIRES spectra, combined Hipparcos and Gaia measurements to probe the astrometric acceleration of Kepler-444 A over 25 years, and archival imaging of the system. With this revision, we could estimate the mass of the binary companion BC without any inference from mass-magnitude relationships. The new fits are consistent with the orbital configuration obtained by \citet{Dupuy2016}. 

Then, we applied the stability-driven refinement technique \citep{Stalport2022} on the dynamical model including the stellar binary companion. We did not observe significant updates in the planet parameters, besides a slight effect on the mass of planets c and f. As expected, the stability-driven approach does not further refine the mass where TTV analyses were able to estimate them -- in that case planets d and e. However, we observe a strong dependency of the BC orbit on the stability of the planetary system. Thanks to this observation, we were able to further tighten the constraints on the BC orbit. A notable result is that the A-BC distance at periastron is estimated to 4.94 AU, with a 68.27$\%$ confidence interval of [3.84, 6.08] AU. Therefore, it is likely that the stellar binary companion passes within 5 AU of the central star. This new result has implications in understanding the conditions with which the planets formed, which appeals for future studies. Notably, the truncation radius of the past protoplanetary disk under the influence of the BC companion can be revised in light of the present conclusion. Indirectly, this might have important consequences on the expected atmospheric composition of these planets, and it will provide insights into the current scenarios. Additionally, the upcoming ESA PLATO mission will provide new valuable TTV measurements that will further constrain the mass of planets d and e, leading to deeper insights into the planets density and supposed composition. Furthermore, these new transit measurements could reveal TTV detection on (some of) the three other planets. The final, updated masses and orbital parameters of the planets and the stellar binary companion BC are presented in Table \ref{tab:Solution}. We encourage future works on Kepler-444 to use those values, which are the most precise up-to-date.  

Finally, we also investigated the dynamical plausibility of a hypothetical sixth outer planet. Using the estimated conjunction time from \citet{Bourrier2017}, we showed that this putative body could exist within a certain range of orbital periods, roughly between 12 and 20 days. Additional targeted observations are needed to shed light on this potential planet. Notably, new measurements with HST in the appropriate temporal window would help to confirm or discard otherwise the existence of the sixth planet.

\begin{acknowledgements}  
We thank the anonymous referee for insightful comments, which significantly improved the presentation of our work. 
This work has been carried out within the framework of the National Centre of Competence in Research PlanetS supported by the Swiss National Science Foundation under grants 51NF40\textunderscore 182901 and 51NF40\textunderscore 205606. The authors acknowledge the financial support of the SNSF.  

This project has received funding from the European Research Council (ERC) under the European Union's Horizon 2020 research and innovation programme (project {\sc Spice Dune}, grant agreement No 947634). 

This work has made use of data from the European Space Agency (ESA) mission Gaia (https://www.cosmos.esa.int/gaia), processed by the Gaia Data Processing and Analysis Consortium (DPAC, https://www.cosmos.esa.int/web/gaia/dpac/consortium). Funding for the DPAC has been provided by national institutions, in particular the institutions participating in the Gaia Multilateral Agreement.

The authors thank Trent Dupuy for helpful conversations.

Tools used: \texttt{REBOUND} \citep{Rein2012} , \texttt{orvara} \citep{orvara}
\end{acknowledgements}

\bibliographystyle{aa} 
\bibliography{bib.bib}

\begin{appendix}
\section{Updated parameters} 
\begin{table}[h!] 
\caption{Updated masses and orbital elements of the planets Kepler-444 Ab, c, d, e, f, and the stellar binary companion Kepler-444 BC.}
\label{tab:Solution}
\centering
\begin{tabular}{@{}lllllll@{}}
\hline \hline
\textit{\underline{Planets}} \\ 
\\ 
Parameter [Units]  & Planet b    & Planet c & Planet d & Planet e & Planet f  \\ 
\hline  \\
$P$ [days]  & 3.600105$\substack{+0.000029 \\ -0.000033}$ & 4.545876(29) & 6.189441$\substack{+0.000056 \\ -0.000037}$ & 7.743451$\substack{+0.000070 \\ -0.000104}$ & 9.740499$\substack{+0.000071 \\ -0.000023}$  \vspace{0.15cm} \\ 
$\lambda$ [deg]  & 0.00$\pm$0.05 & -38.86$\pm$0.04 & 94.04$\pm$0.04 & -167.81$\pm$0.04 & 103.99$\pm$0.01  \vspace{0.15cm} \\ 
$\sqrt{e}\cos(\omega)$  & -0.0284$\substack{+0.1350 \\ -0.1078}$ & 0.0032$\substack{+0.1204 \\ -0.1222}$ & 0.0967$\substack{+0.0643 \\ -0.1220}$ & -0.0374$\substack{+0.1236 \\ -0.0874}$ & -0.0580$\substack{+0.1219 \\ -0.0739}$  \vspace{0.15cm} \\ 
$\sqrt{e}\sin(\omega)$  & 0.0545$\substack{+0.0959 \\ -0.1494}$ & -0.0258$\substack{+0.1320 \\ -0.1074}$ & 0.0270$\substack{+0.1007 \\ -0.0837}$ & 0.0250$\substack{+0.0798 \\ -0.1015}$ & 0.0475$\substack{+0.0754 \\ -0.1153}$  \vspace{0.15cm} \\ 
 $i$ [deg]  & 91.97$\substack{+0.27 \\ -0.31}$ & 92.77$\pm$0.12 & 91.94$\substack{+0.11 \\ -0.10}$ & 90.62$\substack{+0.27 \\ -0.35}$ & 92.08$\pm$0.06  \vspace{0.15cm} \\  
 $\Omega$ [deg]  & 0 (fixed) & 0 (fixed) & 0 (fixed) & 0 (fixed) & 0 (fixed)  \vspace{0.15cm} \\ 
 $m$ [m$_{\oplus}$]  & <0.078 ; <0.111 & <0.146 ; <0.231 & 0.038$\substack{+0.064 \\ -0.022}$ & 0.035$\substack{+0.059 \\ -0.020}$ & <0.286 ; <0.764  \vspace{0.15cm} \\   
 \\ 
\textit{\underline{BC companion}} \\ 
\\ 
Parameter [Units] & & & Index & Index \\ \hline  \\
$a$ [AU] & 39.26$\substack{+0.77 \\ -0.73}$ & & $a$ & Semi-major axis  \vspace{0.15cm} \\ 
$e$ & 0.874$\pm$0.026 & & $e$ & Eccentricity \vspace{0.15cm} \\ 
$\omega$ [deg] & -25.4$\pm$1.7 & & $\omega$ & Arg. of periastron \vspace{0.15cm} \\ 
$\lambda$ [deg] & 177.7$\substack{+2.7 \\ -2.8}$ & & $\lambda$ & Mean long. \vspace{0.15cm} \\ 
$i$ [deg] & 90.89$\substack{+3.78 \\ -3.73}$ & & $i$ & Inclination  \vspace{0.15cm} \\  
$\Omega$ [deg] & 72.9$\substack{+1.4 \\ -1.5}$ & & $\Omega$ & Long. of asc. node \vspace{0.15cm} \\ 
$m$ [m$_{\odot}$]  & 0.633$\substack{+0.019 \\ -0.018}$ & & $m$ & Mass \vspace{0.15cm} \\ 
$a (1-e)$ [AU] & 4.94$\substack{+1.14 \\ -1.10}$ \vspace{0.15cm} \\ \hline
\end{tabular}
\tablefoot{These values take into account the stability constraint. Reported are the median values of the posterior distributions, together with their 68.27$\%$ confidence intervals. For the mass of planets b, c, and f, the values that we report are the upper limits containing 68.27$\%$ and 95.45$\%$ of the posterior.}
\end{table} 
\end{appendix}
\end{document}